%% file: template.tex
\documentclass{Interspeech2024}

\usepackage{subcaption}
\usepackage[table]{xcolor}
\usepackage{multirow}
\usepackage{cleveref} 
\usepackage{todonotes}
\usepackage{colortbl}
\interspeechcameraready

\title{Harder or Different? Understanding Generalization of Audio Deepfake Detection}
\name[affiliation={1}]{Nicolas M.}{Müller}
\name[affiliation={2}]{Nicholas}{Evans}
\name[affiliation={3}]{Hemlata}{Tak}
\name[affiliation={1}]{Philip}{Sperl}
\name[affiliation={1}]{Konstantin}{Böttinger}

\address{
  $^1$Fraunhofer AISEC $^2$EURECOM $^3$Pindrop, USA
}
\email{nicolas.mueller@aisec.fraunhofer.de}


\newcommand{\newpara}[1]{\vspace{4pt}\noindent\textbf{#1}}
\begin{document}

\maketitle

\begin{abstract}
% Text Nick 
Recent research has highlighted a key issue in speech deepfake detection: models trained on one set of deepfakes perform poorly on others. The question arises: is this due to the continuously improving
quality of text-to-speech (TTS) models, i.e., are newer DeepFakes just ‘harder’ to detect? Or, is it because deepfakes generated with one model are fundamentally different to those generated using another model? We
answer this question by decomposing the performance gap between in-domain and out-of-domain test data into ‘hardness’ and ‘difference’ components. Experiments performed using ASVspoof databases indicate that
the hardness component is practically negligible, with the performance gap being attributed primarily to the difference component. This has direct implications for real-world deepfake detection, highlighting that
merely increasing model capacity, the currently-dominant research trend, may not effectively address the generalization challenge.

\end{abstract}

\input{content}

\bibliographystyle{IEEEtran}
\bibliography{mybib}

\end{document}

%% file: content.tex
\section{Introduction}
%@ NICK: Implemented all of your suggestions, and resolved the comments.
% Please do feel free to edit direclty in the text, I have enabled track changes.
Recent years have seen tremendous advances in machine learning (ML).
One area in which progress has been particularly impressive is that of text-to-speech (TTS) synthesis. 
It is now possible to generate high-quality, convincing speech signals which mimic closely the voice identity of specific individuals~\cite{chen2020wavegrad,betker2023better}.
Numerous online services, e.g.~\cite{elevenlabs,Voicemak29:online, googletts}, enable this  technology to be used by \emph{anyone},  even those without any relevant technical expertise.
While it has plentiful legitimate applications, its accessibility has also led to serious threats, e.g.\ fraud, misinformation, and defamation~\cite{audio-scam, audio-scam-2, audio-fake-news,steiswri94:online}. 

A potential solution lies in the ML-driven detection of such deepfakes using, for example, binary classifiers to discriminate between genuine/bonafide and AI-generated speech. 
The field has witnessed a surge in research, from the creation of extensive datasets~\cite{in-the-wild,cfad,had,partial-spoof,fakeavceleb,sing-fake,mlaad}
to the development of new detection models~\cite{wang2021Comparative, muller2023complex,tak2021EndtoEnd,ge2021raw,aasist,whisper-df}.
Most notably, initiatives such as ASVspoof~\cite{asvspoof-2015, asvspoof-2017, asvspoof2021_challenge_and_dataset_desc} which were launched to benchmark competing detection solutions, seemingly show impressive progress; 
lower and lower state-of-the-art error rates are reported on a regular basis~\cite{aasist,whisper-df,liu2023leveraging}. 
However, reliability in real-world scenarios often remains untested, while there are reports that generalisation to out-of-domain scenarios is wanting.
The limited ability to generalize to deepfakes generated using new attack algorithms, or even algorithms that are simply different to those used to create training data, has been and continues to be a source of major concern~\cite{in-the-wild}.
%Detection solutions which fail to generalise can furthermore not be replied upon to function as expected when deployed in the wild.

\begin{figure}[t]
    \centering
    \includegraphics[width=0.5\textwidth]{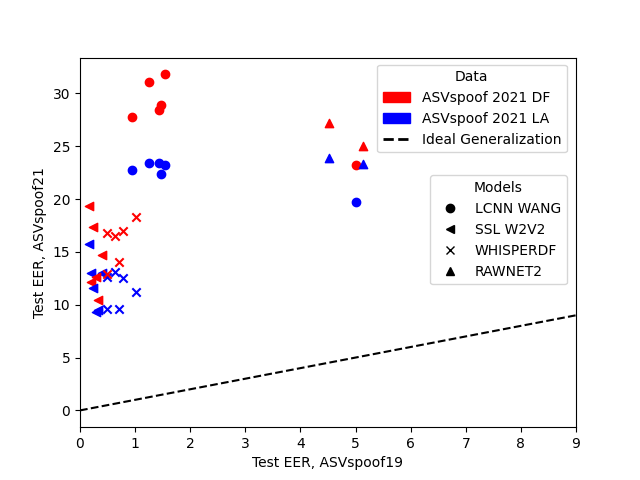}
    \caption{Comparison of EER between ASVspoof 2019 (x-Axis) and ASVspoof 2021 (y-axis) for models trained on ASVspoof 2019. None of the model achieves ideal generalization, as denoted by the dashed line.}
    \label{fig:generalization}
\end{figure}

%To move forward, it's essential to understand why detection models
Despite generalisation having been a focal point of related research for almost a decade, we remain far from practical deepfake detection solutions which generalise to attacks and acoustic conditions seen in the wild, c.f.~\Cref{fig:generalization}.
Key to unlocking progress is an understanding of why current detection solutions fail to generalize. 
The answer can be attributed to two main factors. 
First, for want of a better term, `hardness': the increasing sophistication of speech deepfakes may make them inherently more challenging to detect.
%This hypothesis is supported by the finding that the latest TTS and voice conversion (VC) technology can produce speech signals that are increasingly difficult to detect, not only in the case of human listeners, but also automatic deepfake detection solutions. 
Second, `difference': the characteristics used by a detection model to discriminate between bonafide speech and deepfakes may not generalize across different attack algorithms.
This implies that detection failures might not result from a lack of detection model capacity, but from fundamentally different deepfake characteristics. 
Thus, the issue of generalization may be due to either `hardness' or `difference', or a combination thereof.

\textbf{Contributions.}
We report a means to decompose the performance discrepancy between training and out-of-domain test data % when detecting voice spoofs 
into two components: `hardness' and `difference'. 
We report a study, performed using four different detection models and the ASVspoof~2019 and ASVspoof~2021 datasets, which shows the following.

\begin{itemize}
    \itemsep0em 
    \item When using truncated utterance sub-segments (rather than the full utterance) selected from the ASVspoof 2019 database, performance is substantially diminished, which we show can be attributed mainly to `hardness'.
    \item For the ASVspoof 2021 logical access (LA) dataset, which contains variation related to the use of different compression and telephony encoding algorithms, degraded detection performance can be attributed to both `hardness' and `difference'.
    \item In contrast, for the ASVspoof 2021 deepfake (DF) dataset which contains data collected from multiple sources, degraded detection performance can be attributed almost exclusively to `difference'.
    This observation also holds for the In-the-Wild database~\cite{in-the-wild}.
\end{itemize}
These findings suggest that efforts to extend model capacity, while beneficial in the case of in-domain benchmarking, are insufficient and might even be detrimental to the pursuit of generalisable detection solutions.
We argue that this calls for the re-focusing of research effort to better understand and address the `difference' between deepfakes seen in the wild. 
%These findings suggest that only increasing model capacity, which simply tackles "hardness", may not solve the challenge of model generalization. 
%To achieve reliable improvements, we need new methods to handle and understand the "difference" between individual deepfakes, such as better Explainable AI (XAI).

\section{Related Work}
A substantial volume of research in speech deepfake detection was performed in the scope of ASVspoof challenges~\cite{asvspoof-2015, todisco2019asvspoof, asvspoof-2017, asvspoof2021_challenge_and_dataset_desc}. This body of work established a benchmark which was used in subsequent work as a reference point to assess performance~\cite{wang2021Comparative, muller2023complex,tak2021EndtoEnd,ge2021raw,aasist,whisper-df}. 
However, recent observations highlight that promising results obtained using ASVspoof databases do not necessarily  translate to reliable detection in real-world scenarios~\cite{muller2021Speech, in-the-wild}. This finding has prompted the creation of new datasets designed to address the performance gap~\cite{mlaad}. 
Research has also explored the detection of so-called partial-spoofs, where bonafide utterances are segmented and then concatenated with content generated using text-to-speech synthesis~\cite{partial-spoof}.  Other directions include the investigation of deepfake detection for singing voices~\cite{sing-fake}. 
%While the focus is gradually shifting towards improving generalization 
Despite progress, generalisation~\cite{jain2021improving} and robustness~\cite{barrington2023single} remain in focus. New attacks continue to emerge, hence the challenge is as great as ever.
%a comprehensive analysis of the underlying challenges in generalization of speech deepfake detection remains unaddressed. 
Whereas specific studies of generalisation have been reported in other fields, e.g.\ computer vision and 
%field have attracted attention ,  has more thoroughly examined generalization issues in 
object detection~\cite{recht2019imagenet,lu2020harder}, a similar comprehensive analysis of the underlying challenges in speech deepfake detection is lacking.  The related work in computer vision serves as inspiration and as a methodological foundation for the study reported in this paper.

\section{Methodology}

Let $D$ and $D'$ be speech databases (Section~\ref{s:databases}), each of which is partitioned into a training and test set. Model performance is evaluated using the Equal Error Rate (EER)~\cite{asvspoof2021_challenge_and_dataset_desc,todisco2019asvspoof}. Define in-domain model performance $D_\rightarrow D$ as the EER for the test partition of $D$ after training on the training partition of $D$. Similarly, out-of-domain model performance $D_\rightarrow D'$ denotes the EER when the model is trained using the training partition of $D$ and evaluated using the test partition of $D'$.

To gain insights into the generalization performance of a given model, we analyze the performance gap:

\begin{align}
D_\rightarrow D - D_\rightarrow D',
\label{eq:1}
\end{align}

\noindent which represents the gap in detection performance for in-domain $D_\rightarrow D$ and out-of-domain $D_\rightarrow D'$ scenarios. It can be decomposed into a `hardness' and 'difference' components:

\begin{equation}
\begin{aligned}
& \underbrace{D _\rightarrow D - D _\rightarrow D'}_{\text{Performance gap}} \\
=\quad & \underbrace{D _\rightarrow D - D' _\rightarrow D'}_{\text{Hardness gap}} + \underbrace{D' _\rightarrow D' - D _\rightarrow D'}_{\text{Difference gap}}
\end{aligned}
\label{eq:2}
\end{equation}

\noindent These components can be interpreted as follows:
\begin{itemize}
\itemsep0em
\item The hardness gap is a measure of the relative difficulty in terms of the  in-domain detection performance $D_\rightarrow D$ and $D'_\rightarrow D'$.
\item The difference gap reflects the difference in detection performance when a model trained using database $D$ is then tested using database $D'$. % from the one it was trained on, $D$.
\end{itemize}

\noindent When analyzing the performance gap using \Cref{eq:2}, two outcomes are possible. First, a larger `hardness gap' indicates that $D'$ is inherently more difficult than $D$. This difficulty may arise from the presence of higher-quality deepfakes in $D'$ which contain fewer artefacts and which are hence  
%It is reasonable to assume that higher-quality spoofs are 
harder to detect.

A second possible outcome is a larger `difference gap', which suggests that the anti-spoofing model does not generalize well from $D$ to $D'$, even though $D'$ is not inherently more challenging. Put differently, if the `difference gap' is large, then the `hardness gap' must be small, which means that $D_\rightarrow D$ and $D'\rightarrow D'$ are equally challenging. 
Thus, poor results in $D\rightarrow D'$ are not related to the complexity of $D'$, and more the result of the model being overfitted to $D$. Learned artefacts may even stem from shortcuts, as described in~\cite{muller2021Speech}, or other features which, while highly informative for $D$, do not transfer to $D'$.

By analyzing and decomposing the performance gap as described above, we can gain insights into whether poor performance in $D_\rightarrow D'$ is simply the result of $D'$ being more challenging, or whether the two data distributions $D$ and $D'$ are simply too different for the model to generalize.

% If the performance gap mainly stems from from inherently more challenging test data, such as more sophisticated deepfakes,
% then we anticipate a more substantial `hardness gap'. 
% On the other hand, 
% if deepfakes contained in the test data exhibit a similar level of quality but differ in acoustic conditions 
% compared to the training data, then the performance gap should be attributed to a greater `difference' gap.
% We expect the same in the case that deepfakes contained in the test data are no more `hard' to detect, if only learning was performed using similar deepfakes among the training data, hence the distinction between what is `hard' and what is `different'. 
% If the "performance gap" is primarily due to "difference", then enhancing model capabilities may not address the underlying issue.

\section{Evaluation}
We report an evaluation performed using four different detection models:
an LCNN model~\cite{wang2021Comparative}; RawNet2~\cite{tak2021end};  Whisper-DF~\cite{whisper-df}; SSL-W2V2~\cite{ssl_antispoof}.
These models represent the most common approaches to speech deepfake detection, and were state-of-the-art at the time of their publication.
All models were trained using a batch size of 16, the Adam optimizer with a learning rate of $10^{-3}$, and for up to $500$ epochs.
To prevent overfitting, we use aggressive early stopping with a hyper-parameters $\delta = 5 \cdot 10^{-3}$ train EER and a patience of $1$ epoch.
Additionally, we remove leading and trailing silence from all data, to avoid the use of non-speech shortcuts reported in~\cite{muller2021Speech}.

\subsection{Databases}\label{s:databases}
%\begin{itemize}
Experiments were performed using the following three databases, each of which is partitioned into training (80\%) and test (20\%) sets.

\newpara{ASVspoof 2019 LA~\cite{todisco2019asvspoof}}: We use ASVspoof 2019 LA for both training and in-domain testing. It has 
%is divided into 
three disjoint partitions: training, development and evaluation. Importantly, the evaluation partition includes out-of-domain data, namely, attacks not seen during training. Since our focus is on using ASVspoof 2019 LA for training and in-domain testing, we disregard these partitions and reorganize the entire dataset into an $80\%$ training and $20\%$ in-domain testing split.

\newpara{ASVspoof 2021~\cite{yamagishi2021asvspoof}}: Out-of-domain evaluation was performed using ASVspoof 2021 databases, specifically the `Logical Access' (LA) dataset and the `Deepfake' (DF) dataset. 
ASVspoof 2021 LA data contains simulated channel variability, with all audio files being processed through real telephony systems, including voice-over-internet-protocol (VoIP) systems and public switched telephone networks (PSTN)~\cite{asvspoof2021_challenge_and_dataset_desc}. 
Consequently, detection models must handle compression artifacts, packet loss, and other related issues. 
On the other hand, the ASVspoof 2021 DF dataset contains deepfakes sourced from the 2018 and 2020 Voice Conversion Challenge databases~\cite{lorenzo2018voice, zhao2020voice}. 
A subset of the DF dataset also contains additional variability stemming from the application of various lossy media storage codecs.
%for media storage are used  to mimic realistic scenarios.

\newpara{In-the-Wild}~\cite{in-the-wild}: This dataset includes both synthesized and genuine audio for 58 politicians and public figures, obtained from social networks and video platforms. 
It encompasses 20.8 hours of authentic and 17.2 hours of fake audio. 
Models trained using ASVspoof 2019 have been shown to generalise poorly to in-the-wild data~\cite{in-the-wild}.
%\end{itemize}

\subsection{Experiments}
Our objective is to assess the decomposition of the performance gap into `hardness' and `difference' components across various scenarios.
%\begin{itemize}

\newpara{Audio Degradation} -- In our initial experiment, we deliberately lower the quality of the test data by aggressively truncating utterance duration.
In this scenario, with respect to \Cref{eq:2}, $D$ denotes the original ASVspoof 2019 dataset with $8$ seconds (s) of input, while $D'$ represents the same data where each sample is randomly truncated to $0.25s$. 
For models that require a minimum audio length (such as WhisperDF and SSL-W2V2), the truncated input is
appropriately concatenated after truncation in order to produce utterances of the required duration, c.f. \Cref{sec:result_audio_deg}.

\newpara{Unseen databases} -- The second, third, and fourth experiments evaluate the performance gap between ASVspoof 2019 LA and other databases: 
ASVspoof 2021 LA, DF, and In-the-Wild. 
% These scenarios represent progressively increasing complexity levels. 
The ASVspoof 2021 LA database includes the same attacks\footnote{In this context, an `attack' refers to a distinct Text-to-Speech (TTS) synthesis algorithm. Introducing a new and unseen attack thus involves the inclusion of a new and previously unencountered TTS synthesis algorithm.} as the ASVspoof 2019 database, but incorporates lossy compression and telephone encodings.  The ASVspoof 2021 DF database includes a broader variety of attacks sourced from the voice conversion challenge databases~\cite{lorenzo2018voice, zhao2020voice}. 
Lastly, the In-the-Wild dataset consists of 
unknown attacks which likely include some which are different to those in the ASVspoof databases.
% entirely uncontrolled,  and potentially high-quality attacks due to its recency compared to ASVspoof 2019. 
The results of these experiments are presented in \Cref{ss:results_21la} and \Cref{ss:results_21df_itw}.

\section{Results}
\subsection{Experiment 1: Audio Degradation}\label{sec:result_audio_deg}
\begin{table}[t]
    \centering
     \caption{Decomposition of the performance gap into `hardness' and `difference' for $D$ = ASVspoof 2019, and $D'$ = ASVspoof 2019 with input length truncated to 0.25s. 
    A substantial portion of the performance gap can be attributed to `hardness' (blue).
    Surprisingly, for some models, there also is a rather substantial contribution from `difference' (red).
    Results are presented as mean EER ± standard deviation across five individual trials.
    }
    % \resizebox{.48\textwidth}{!}{
        \input{res/tab/ex2/feature_subselection}
    % }
    % \caption{Feature subselection varied, \url{/opt/mueller/output_hardness/feb_26_feat_subselection/}
    % \caption{Decomposition of the performance gap into `hardness' and `difference' for the ASVspoof 2019 dataset, when input length is truncated for test data. 
    % A substantial portion of the performance gap can be attributed to `hardness' (blue).
    % Surprisingly however, for some models, there also is a rather substantial contribution from `difference' (red).
    % Results presented as mean EER ± standard deviation across five individual trials.
    % }
    \label{tab:asv19_degradation}
\end{table}

Recall that as per~\Cref{eq:2},  the `performance gap' is the sum of the `hardness gap' and `difference gap'.~\Cref{tab:asv19_degradation} illustrates this breakdown of performance degradation due to reduced input length.
We observe that the main reason for this is the `hardness gap', as marked in blue. 
As expected, diminishing the amount of information available to the model inherently increases the problem `hardness'. 
The most pronounced effect is seen for the RawNet2 model:
a $33.4\%$ increase in EER, of which $30.4\%$ EER can be attributed to increased `hardness', while only $3.0\%$ EER can be attributed to the `difference' in conditions. 
Similarly, performance for the LCNN model deteriorates by $17\%$ EER, of which $13\%$ EER pertains to an increase in `hardness'.
Larger, pre-trained models like SSL W2V2 and WhisperDF, exhibit less sensitivity to the `hardness gap', but are still substantially affected.
Interestingly, there is also a substantial contribution from the `difference'.

\subsection{Experiment 2: ASVspoof 2021 LA}\label{ss:results_21la}

\Cref{tab:asv19vs21la} details results for the ASVspoof 2021 LA dataset.
%, which contains the same audio files as ASVspoof 2019, but which have been altered substantially through the use of encoding and compression techniques. Surprisingly, contrary to expectations these modifications would increase the dataset's 
Even if the differences in compression and encoding are expected to compound the challenge, our experiments indicate otherwise: 
although performance deterioration is less pronounced than for input data reduction, the primary factor for the performance gap is attributed to `difference'. Notably, for the LCNN model, the smallest among the four analyzed, the performance gap can also be attributed to increased hardness. 
For the other, larger models, it seems that the presence of compression and encoding variation does not increase hardness.

\begin{table}[t]
    \centering
    \caption{Decomposition of the performance gap where $D$ = ASVspoof 2019 and $D'$ = ASVspoof 2021 LA.
    For most models, the performance gap can be attributed solely to `difference' (red).
    }
    % \resizebox{.48\textwidth}{!}{
        \input{res/tab/asv21la}
    % }
    % \caption{Decomposition of the performance gap when training on ASVspoof 2019 and evaluating on ASVspoof 2021 LA.
    % For most models, the performance gap can be attributed solely to `difference' (red).
    % }
    \label{tab:asv19vs21la}
\end{table}

\subsection{Experiment 3, 4: ASVspoof 2021 DF \& In-The-Wild}\label{ss:results_21df_itw}

\begin{table}[t]
\centering
\caption{Decomposition of performance gap between $D$ = ASVspoof 2019 and $D'$ = ASVspoof 2021 DF.
The performance gap can be attributed solely to `difference' (red).
}
% \resizebox{.48\textwidth}{!}{
        \input{res/tab/asv21df}
    % }
% \caption{Decomposition of performance gap between ASVspoof 2019 and ASVspoof 2021 DF.
% The performance gap can be attributed solely to `difference' (red).
% }
\label{tab:asv19vs21df}
\end{table}

\begin{table}[t]
    \centering
     \caption{Decomposition of performance gap between $D$ = ASVspoof 2019 and $D'$ = In-the-Wild. The performance gap can be attributed solely to `difference' (red).}
    % \resizebox{.48\textwidth}{!}{
        \input{res/tab/itw}
    % }
    % \caption{Decomposition of performance gap between ASVspoof 2019 and In-the-Wild. The performance gap can be attributed solely to `difference' (red).}
    \label{tab:asv19vsITW}
\end{table}

The final experiment assesses the trade-off between hardness and difference for unseen attacks. 
For this purpose, 
testing is performed using the ASVspoof 2021 DF and the In-the-Wild databases.
% ASVspoof 2021 DF, containing partially unseen attacks, and In-the-Wild, encompassing completely unseen and more recent attacks, are utilized.
\Cref{tab:asv19vs21df} displays results for the ASVspoof 2021 DF database, for which an average performance gap of $14\%$ to $26\%$ in EER is observed, and predominantly attributed to the `difference gap'. 
Results for the In-the-Wild database, c.f. \Cref{tab:asv19vsITW}, reveal an even more substantial performance gap ($31\%$ to $78\%$), which is entirely due to the `difference gap'.
Consequently, all models perceive the new attacks as a domain shift, characterized not by an escalation in difficulty or complexity but rather as 
an exposure to previously unseen, distinct attacks with different characteristics.
% encountering new, unseen, and distinct data and/or features.

\section{Discussion and Implications}\label{s:discus}

% The experiments described in the previous sections reveal the following insights:
% Detection models, including state-of-the-art ones like SSL-W2V2, struggle to adapt to changes in input data:
% both to simple modifications, such as input length adjustments, as well as to more intricate data alterations, such as encoding and lossy compression seen in ASVspoof 2021 LA, as well as entirely new data encountered in ASVspoof 2021 DF and In-the-Wild.
% We can analyze this inability to generalize through a breakdown of the performance gap into two components: `hardness' and `difference'. 
% In the extreme cases, this decomposition aligns with what one would expect:
% Severe input length truncates makes the model input more noisy and thus, the problem becomes `harder'.
% Similarly, we expected completely new and unseen attacks to be - at least to some degree  - `different', which is indeed what we observe.

%Notwithstanding, t
% These results are astounding.
Results reported above are perhaps surprising.
%Initially, we had expected data augmentation to increase `hardness'. %, not `difference'.
We expected both input length truncation and variability in compression and encoding to increase `hardness' 
since the introduction of nuisance variability to the test data  usually increases classification difficulty.
%could be considered the archetype of increasing data difficulty.
Instead, results show 
%However, for both input length truncation, as well as compression and telephone encodings, we observe 
a marked increase in `difference'.
%To us, this is very surprising, since introducing more noise to the test data 
%could be considered the archetype of increasing data difficulty.

Second, whenever models are presented with new data, i.e. unseen attacks, the lack of generalization performance is also attributed near-exclusively to `difference'.
This is in stark contrast to similar experiments in the vision domain, where Lu et al.\ evaluate the generalization of models trained on CIFAR-10 to the out-of-domain database CIFAR-10.2~\cite{lu2020harder}.
Here, a `performance gap' of about $15\%$ accuracy is, on average, decomposed into a `difference gap' of $5\%$ and a `hardness gap' of $10\%$.
This more balanced decomposition is different from our experiments on ASVspoof 2021 DF and In-the-Wild, where the `difference gap' is the sole contributor to performance degradation.

Solving `harder' databases might become feasible with the development of more advanced models~\cite{lu2020harder}.
However, `difference' poses a more elusive challenge and may not be effectively addressed by merely increasing model capacity. 
Recall that machine learning is fundamentally pattern recognition. 
Patterns falling outside the training data are inherently difficult for a model to comprehend, as the model's knowledge is derived entirely from the statistical aggregation of input-output relationships in the training data. 
Consequently, the consistent prevalence of the `difference' gap across all experiments, particularly as the dominant factor for ASVspoof 2021 and In-the-Wild, may explain why achieving true out-of-domain generalization has been so problematic in previous research. 
It appears that deepfake detection and anti-spoofing models are overly specific to their training data, with even minor deviations being treated as outside the training data distribution (i.e., `different').

% Opposed to overfitting on the training \emph{data}, models which exhibit \emph{distribution}-overfitting do not remember individual samples, but 
We induce that models learn specific features that perform well for the current database, but that are so specific that they don't transfer well to other out-of-domain databases.
An example from related work is the `silence-shortcut' in the ASVspoof 2019 dataset~\cite{muller2021Speech}:
Here, models can use the length of leading silence to achieve near-perfect performance.
The length of the silence correlates with the class label; and even generalizes to the ASVspoof 2019 test data, but is obviously not a semantically meaningful feature that can be expected to work in the real world.
It is likely that there are other such `shortcuts'~\cite{geirhos2020shortcut} in voice anti-spoofing datasets, which may be harder for humans to identify, but which nevertheless are very predictive clues that are exploited by anti-spoofing models.
The existence of such 
 `shortcuts' would align with the dominance of the `difference gap', which is not nearly as pronounced in, for example, the vision domain~\cite{lu2020harder}.
%, highlighting a critical limitation in their ability to adapt to novel scenarios.

Therefore, while increasing model capacity by adding more parameters may prove beneficial in other domains, such as generating text with large language models, it might not be as effective in the realm of supervised classification for audio anti-spoofing and deepfake detection. 
%TODO leave this in?
% In these areas, models may tend to overfit to the training data distribution. 
% In the worst case, more parameters may even exasperate the overfitting.
At the same time, the use of self-supervised learning (SSL), exemplified by SSL-W2V2 and WhisperDF, appears to be a promising approach, as indicated by the overall smaller `performance' gaps.
Possibly, such pre-trained models may extract less specific, and thereby better generalizing features.
%This might help them to extract less specific, and thus better generalizing features.
%This might make them more resilient against the overfitting issues previously mentioned.

\section{Future Work and Conclusion}
We present in this paper a study which demonstrates that the performance gap in anti-spoofing generalization can be decomposed into `hardness' and `difference' components. 
Our hypothesis is that poor generalization is due mainly to the substantial yet inadequately addressed influence of the `difference' gap, indicating that detection models overly adapt to the training \emph{distribution}.

Future research might %should aim to deepen the understanding of this issue by 
systematically analyze and break down errors across a wide range of audio augmentations (such as, potentially, augmentation by band-pass filters, compression, room impulse responses, Gaussian noise, time- and pitch-shifting, etc.), as well as between different attacks, e.g. between autoregressive and transformer-based TTS models. 
Such investigations could provide a more comprehensive view of the conditions under which models generalize well instead of learning artefacts which are too specific to attacks observed in training data.
%versus those in which they latch onto features highly specific to the training dataset.
Such research may be more beneficial than the current concentration upon increasing model capacity, a strategy which may address only the `hardness', but ignore the key `difference' problem in speech spoofing and deepfake detection.

% \begin{table*}[]
%     \centering
%     \resizebox{.999\textwidth}{!}{
%         \input{res/tab/ex2/audiomentations}
%     }
%     \caption{Audiomentations varied, \url{/opt/mueller/output_hardness/feb_22_audiomentations}}
%     \label{tab:my_label}
% \end{table*}

% \begin{table*}[]
%     \centering
%     \resizebox{.999\textwidth}{!}{
%         \input{res/tab/ex2/background_music_or_noise}
%     }
%     \caption{Background Music or Noise varied, \url{/opt/mueller/output_hardness/feb_22_add_background_music_or_noise/}}
%     \label{tab:my_label}
% \end{table*}

%% file: res/tab/ex2/feature_subselection.tex
\begin{tabular}{l|lll}
\toprule
% Model&\multicolumn{2}{c|}{Input (s)}&Perform. & Hardn. & Diff. \\
 % &          Train  &                 Test &      Gap & Gap & Gap \\
Model & Performance & Hardness & Difference \\
 & Gap & Gap & Gap \\
\midrule
 LCNN &     $17.7 \pm 0$ & \cellcolor{blue!25} $13.4 \pm 2$ &    $4.2 \pm 2$ \\
 %  & 8.0&  0.5 & $13.9 \pm 1.3$ &    $12.2 \pm 1$ &  $7.1 \pm 1$ &    $5.1 \pm 2$ \\
 %  & 8.0&    1 &  $8.8 \pm 1.8$ &     $7.1 \pm 2$ &  $4.6 \pm 1$ &    $2.6 \pm 2$ \\
  % & 8.0&    2 &  $4.2 \pm 0.2$ &     $2.5 \pm 0$ &  $1.3 \pm 1$ &    $1.1 \pm 0$ \\
 %  & 8.0&    4 &  $2.0 \pm 0.4$ &     $0.3 \pm 1$ &  $0.1 \pm 0$ &    $0.3 \pm 0$ \\
%   & 8.0&    8.0&  $1.9 \pm 0.5$ &     $0.2 \pm 1$ & $-0.0 \pm 1$ &    $0.2 \pm 1$ \\
% \midrule
   RawNet2 &     $33.4 \pm 4$ & \cellcolor{blue!25} $30.4 \pm 3$ &    $3.0 \pm 5$ \\
   %  & 8.0&  0.5 & $33.4 \pm 0.0$ &    $27.3 \pm 0$ & $23.9 \pm 1$ &    $3.4 \pm 1$ \\
   %  & 8.0&    1 & $27.8 \pm 2.7$ &    $21.7 \pm 3$ & $19.6 \pm 1$ &    $2.1 \pm 3$ \\
    % & 8.0&    2 & $17.6 \pm 6.8$ &    $11.5 \pm 7$ &  $5.4 \pm 0$ &    $6.2 \pm 7$ \\
   %  & 8.0&    4 &  $6.9 \pm 1.0$ &     $0.8 \pm 1$ & $-1.2 \pm 0$ &    $2.0 \pm 1$ \\
%     & 8.0&    8.0&  $6.1 \pm 0.3$ &    $-0.0 \pm 0$ & $-0.0 \pm 0$ &   $-0.0 \pm 0$ \\
% \midrule
  SSL W2V2 &     $23.1 \pm 2$ & \cellcolor{blue!25} $11.8 \pm 2$ & \cellcolor{red!25}  $11.2 \pm 3$ \\
  %  & 8.0&  0.5 & $15.7 \pm 1.6$ &    $15.5 \pm 2$ &  $7.1 \pm 1$ &    $8.4 \pm 2$ \\
  %  & 8.0&    1 &  $7.7 \pm 0.8$ &     $7.5 \pm 1$ &  $3.7 \pm 1$ &    $3.8 \pm 1$ \\
   % & 8.0&    2 &  $1.6 \pm 0.1$ &     $1.5 \pm 0$ &  $0.7 \pm 0$ &    $0.7 \pm 0$ \\
  %  & 8.0&    4 &  $0.2 \pm 0.2$ &     $0.1 \pm 0$ &  $0.0 \pm 0$ &    $0.1 \pm 0$ \\
%    & 8.0&    8.0&  $0.2 \pm 0.1$ &     $0.0 \pm 0$ & $-0.0 \pm 0$ &    $0.0 \pm 0$ \\
% \midrule
 WhisperDF &     $25.1 \pm 1$ &  \cellcolor{blue!25} $8.4 \pm 1$ & \cellcolor{red!25}  $16.6 \pm 1$ \\
 %  & 8.0&  0.5 & $18.6 \pm 1.2$ &    $17.8 \pm 1$ &  $5.4 \pm 1$ &   $12.4 \pm 1$ \\
 %  & 8.0&    1 & $11.4 \pm 1.6$ &    $10.6 \pm 2$ &  $3.1 \pm 1$ &    $7.6 \pm 2$ \\
  % & 8.0&    2 &  $3.8 \pm 0.3$ &     $3.0 \pm 1$ &  $2.6 \pm 1$ &    $0.5 \pm 0$ \\
 %  & 8.0&    4 &  $1.0 \pm 0.5$ &     $0.3 \pm 1$ &  $0.2 \pm 1$ &    $0.1 \pm 1$ \\
  % & 8.0&    8.0&  $0.7 \pm 0.5$ &    $-0.0 \pm 1$ & $-0.0 \pm 1$ &   $-0.0 \pm 1$ \\
\bottomrule
\end{tabular}

%% file: res/tab/asv21la.tex
\begin{tabular}{l|lll}
\toprule
Model & Performance & Hardness & Difference \\
 & Gap & Gap & Gap \\
\midrule
LCNN &    $20.5 \pm 2$ & \cellcolor{blue!25} $7.2 \pm 3$ &   \cellcolor{red!25} $13.4 \pm 3$ \\
RawNet2 &    $15.3 \pm 7$ & $-0.6 \pm 7$ &  \cellcolor{red!25} $15.9 \pm 1$ \\
SSL W2V2 &    $11.8 \pm 2$ &  $0.6 \pm 0$ &   \cellcolor{red!25}$11.2 \pm 2$ \\
WhisperDF &    $10.7 \pm 2$ &  $0.8 \pm 0$ &  \cellcolor{red!25}  $9.9 \pm 2$ \\
\bottomrule
\end{tabular}

%% file: res/tab/asv21df.tex
\begin{tabular}{l|lll}
\toprule
Model & Performance & Hardness & Difference \\
 & Gap & Gap & Gap \\
\midrule
LCNN &    $26.6 \pm 3$ &  $1.7 \pm 2$ &  \cellcolor{red!25} $25.0 \pm 3$ \\
RawNet2 &    $18.1 \pm 7$ & $-2.2 \pm 7$ & \cellcolor{red!25}  $20.3 \pm 2$ \\
SSL W2V2 &    $14.2 \pm 3$ &  $0.1 \pm 0$ &  \cellcolor{red!25} $14.0 \pm 3$ \\
WhisperDF &    $15.2 \pm 2$ &  $0.1 \pm 0$ &   \cellcolor{red!25}$15.1 \pm 2$ \\
\bottomrule
\end{tabular}

%% file: res/tab/itw.tex
\begin{tabular}{l|lll}
\toprule
Model & Performance & Hardness & Difference \\
 & Gap & Gap & Gap \\
\midrule
LCNN &   $78.2 \pm 15$ & $-1.5 \pm 2$ & \cellcolor{red!25} $79.7 \pm 15$ \\
RawNet2 &    $40.6 \pm 7$ & $-6.4 \pm 7$ & \cellcolor{red!25}   $47.0 \pm 3$ \\
SSL W2V2 &    $30.3 \pm 5$ &  $0.1 \pm 0$ &  \cellcolor{red!25} $30.2 \pm 5$ \\
WhiserDF &    $31.4 \pm 5$ & $-0.5 \pm 0$ &   \cellcolor{red!25}$31.9 \pm 5$ \\
\bottomrule
\end{tabular}